\documentclass[10pt,letterpaper]{article}
\usepackage[top=0.85in,left=2.75in,footskip=0.75in]{geometry}

\usepackage{amsmath,amssymb}

\usepackage{changepage}

\usepackage{textcomp,marvosym}

\usepackage{cite}

\usepackage{nameref,hyperref}

\usepackage[right]{lineno}

\usepackage[nopatch=eqnum]{microtype}
\DisableLigatures[f]{encoding = *, family = * }

\usepackage[table]{xcolor}

\usepackage{array}

\newcolumntype{+}{!{\vrule width 2pt}}

\newlength\savedwidth

\newcommand\thickhline{\noalign{\global\savedwidth\arrayrulewidth\global\arrayrulewidth 2pt}%
\hline
\noalign{\global\arrayrulewidth\savedwidth}}

\raggedright
\usepackage[aboveskip=1pt,labelfont=bf,labelsep=period,justification=raggedright,singlelinecheck=off]{caption}

\makeatletter
\renewcommand{\@biblabel}[1]{\quad#1.}
\makeatother

\usepackage{lastpage,fancyhdr,graphicx}
\usepackage{epstopdf}
\fancyheadoffset[L]{2.25in}
\fancyfootoffset[L]{2.25in}

\begin{document}
\vspace*{0.2in}

\begin{flushleft}
{\Large
\textbf{Regime change in a model for political opinion dynamics} 
}
\newline
\\
Alexander Jochim\textsuperscript{1*}, 
Stefan Bornholdt\textsuperscript{1, 2}
\\
\bigskip
\textsuperscript{1} Institute for Theoretical Physics, University of Bremen, Germany\\
\textsuperscript{2} Complexity Science Hub, Vienna, Austria
\\
\bigskip

%
%





* jochim@itp.uni-bremen.de

\end{flushleft}

\section*{Abstract}
Political crises involving a series of violent events are the rule rather than the exception in history. Here, we use a minimal opinion dynamics model to describe regime change as a contested transition between dominant shared ideas in a society. New opinions spread with a cumulative advantage, and agents do not return to opinions they held before, resulting in recurrent boom-and-bust dynamics without tuned parameters. During a transition, opinion changes grow approximately exponentially and then saturate in the finite system, declining approximately linearly as spreading is limited by the shrinking interfaces between opinion groups. We compare qualitative features of these transitions with historical data of political violence, used as an empirical proxy for contested transitions. A permutation test shows that violent events cluster in time in the dense Chinese and US records. Around their peaks, violent events rise in an accelerating way and then decline; in the Chinese data, the decline is consistent with the approximately linear decline in the model. We also use a fragmentation measure that serves as an early warning signal for transitions in the model. Overall, the patterns in the empirical data and the dynamical behavior of the model could contribute to a better understanding of long-term historical processes.

\section*{Introduction}
Humanity has faced recurrent times of peace and chaos over approximately 10\thinspace000 years of human civilization. A prominent phenomenon is societal collapse, usually defined as a rapid decline of social complexity in only a few decades \cite{tainter1988collapse}. Various theories have been proposed, with a recent review given by \cite{brozovic2023societal}. In modern approaches, society is viewed as a complex system from a more data-driven perspective. The use of large data sets has led to the emerging field of cliodynamics \cite{turchin2008arise,turchin2018historical} involving mathematical modeling of history.

Most often societies in history have not collapsed in the sense of total extinction but rather have encountered a fundamental political change, often accompanied by violent uprisings. A recent cliodynamics approach, the Structural-Demographic Theory \cite{goldstone1991revolution,turchin2009secular,turchin2016ages}, relies on empirical data to model these phenomena, based on the analysis of statistical patterns in the data. Stable and unstable periods in history are often found to be recurrent on the scale of a few centuries. This is further supported by recent additions of large datasets and quantitative models for neolithic \cite{kondor2023explaining}, pre-industrial \cite{turchin2009secular}, as well as modern societies \cite{turchin2016ages}. 

\subsection*{Historical data of political violence events}
\begin{figure}[!h]
\includegraphics[width=1\linewidth]{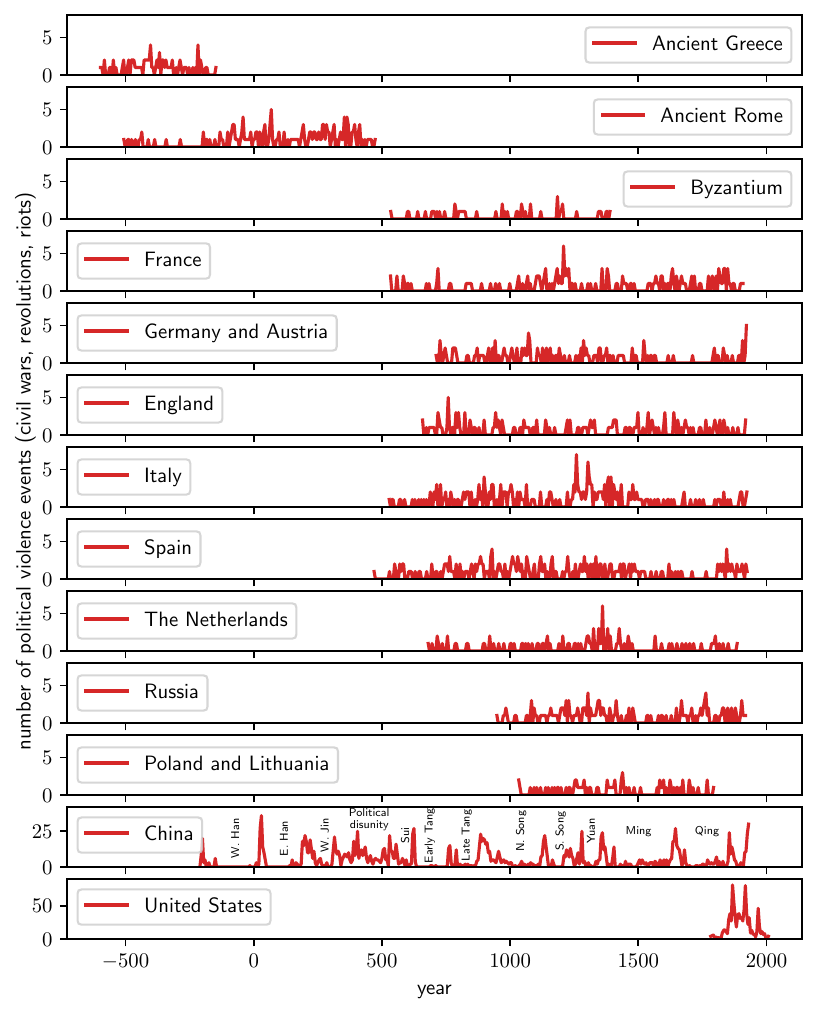} 
\caption{\textbf{Historical data of political violence events.} Number of recorded events for various countries, binned into 5-year intervals. Data for European countries from~\cite{sorokin1937social} (internal disturbances), for China from~\cite{lee1931periodic} (internecine wars) and for the United States from~\cite{turchin2012dynamics} (lethal riots). See Materials and methods for definitions. Dynasties for the Chinese data are included as in~\cite{lee1931periodic}. }
\label{fig:empirical_data}
\end{figure}

Fig~\ref{fig:empirical_data} shows the number of group-level political violence events and internal disturbances within a state, such as internal wars, revolutions and riots, for various countries (see Materials and methods for the definitions used by each source). Periods with many events alternate with periods with few or none. We quantify this in the Results (Table~\ref{tab:clustering}): in the dense Chinese and US data, events cluster on timescales of decades, whereas in the sparser European data clustering is weak or absent.

Another interesting phenomenon that can be observed is that the clustering of events seems to be of similar maximal magnitude for a given country, as seen by peaks of similar magnitude in Fig~\ref{fig:empirical_data}. This is especially true for the Chinese and US data, which are less sparse than the European data. The behavior seems to hold even for millennia as, for example, in the Chinese data. 

One can distinguish historical transitions between the peaks, such as the transition from the Ancient Roman Republic to the Roman Empire or the different dynasties in the historical Chinese Empire, as done in \cite{lee1931periodic}. Old regimes are continuously replaced by new ones. Memory of the past can prevent outdated ones from returning. In a stable phase some common idea (or identity/norm) is shared among society, a concept which is explicitly or implicitly used in various literature on societal collapse such as \cite{tainter1988collapse} and \cite{turchin2009secular}. 

By regime change we mean the replacement of a dominant shared idea in a society, such as a ruling elite, a dynasty or a political order, by a new one. Such transitions are typically contested: competing ideas confront each other until one prevails. Political violence is one observable proxy for this contest, but not the same thing: regimes have changed without large-scale violence, and violence has occurred without a change of regime. We therefore use violent events as a proxy to compare qualitative features of contested transitions with the model, not as a direct measure of regime change.

\subsection*{Mathematical modeling: opinion dynamics and minimal models} 
Mathematically, one can model this common idea as an opinion. In this article, we will use an opinion dynamics approach to modeling the patterns of shifts in shared ideas and phases of societal discord, and subsequently relate the model dynamics to the observed patterns of political violence in the empirical data. Note that there is a difference between the opinion of an individual and the shared ideas in a society as a whole. Groups act as an important structural layer, which can compete in non-violent exchange of views and ideas. However, competition can also occur violently as shown in the empirical data of group-level conflicts within a state in Fig~\ref{fig:empirical_data}. These are the mechanisms that we aim to represent in an agent-based mathematical model of opinion dynamics. 

What spreads in the model can be interpreted in terms of elites, which is motivated in the following. The important role of elites in a society dates back to classical elite theory~\cite{michels1915political,mosca1923elementi,pareto1935mind} and is supported empirically~\cite{schattschneider1975semisovereign,acemoglu2013nations,gilens2014testing}. Turchin and colleagues argue that intra-elite conflict is an essential factor for major political transitions and violent conflict~\cite{goldstone1991revolution,turchin2009secular}. During stable periods, a large support for a specific elite group dominates the political landscape, and a common idea is shared among society. During unstable periods, different elite groups are in conflict with each other, challenging older ideas with potentially violent uprising. A political opinion in this sense is much more dividing than, for example, the preference for one of several democratic parties, and the relevant timescales are inter-generational rather than electoral. Opinion spreading can therefore be read as elites being replaced by new elites. Mathematically, it is the same in the model whether an agent changes its opinion or is replaced by a new agent holding that opinion, and replacement may well be the dominant process on long inter-generational time scales: fundamental political views develop in the early years of an individual and are mostly not altered afterwards in their professional life, known as the impressionable years hypothesis~\cite{newcomb1943personality,krosnick1989aging}.

Models for opinion dynamics have been successful in understanding underlying mechanisms, as they connect individual interactions and macroscopic patterns. In recent decades, a rich variety of models have been developed. Early pioneering work brought classical models such as the voter model \cite{clifford1973model}, models for segregation \cite{sakoda1949minidoka,schelling1971dynamic,hegselmann2017sakoda} and cultural dissemination \cite{axelrod1997dissemination}. Influential later models include the Sznajd model \cite{sznajd2000opinion}, the Majority rule model \cite{galam2002minority} and bounded confidence models \cite{deffuant2000mixing,hegselmann2002opinion}. Reviews of the current state of such modeling approaches are given by \cite{castellano2009statistical,starnini2025opinion} who take the fruitful perspective of statistical mechanics and dynamical systems.

Some models in opinion dynamics have in fact looked at possible mechanisms for how new replaces the old, as for example in modeling  the eternal change of trends in fashion \cite{donangelo2000physics}, but also in the case of scientific paradigms replacing each other \cite{bornholdt2011emergence}, quite similar to the dynamics in epidemics where new viruses replace the old due to immunity that accumulates in a population of hosts \cite{sneppen2010minimal}. In this sense, the model used here is closely related to susceptible-infected models (SI models) with acquired immunity: an agent that has once held an opinion is ``immune'' to it in the future. It is also related to models for the diffusion of innovations, where the cumulative number of adopters follows a characteristic S-shaped curve~\cite{rogers2003,bass1969}. In well-mixed populations, adoption slows down exponentially as the remaining non-adopters become rare, whereas in spatially local interaction the adoptions are approximately proportional to the interface length between opinion groups. We return to this distinction in the Results. Here we will use that model idea for how old shared ideas in society are replaced by new ones, and ask which qualitative features of contested transitions it reproduces.

Let us define an opinion dynamics model along these lines and keep it as simple as possible. Minimal models have proven a useful tool in isolating and studying mechanisms at work in complex systems \cite{marsili2024simplicity}, and may provide a complementary tool to study patterns in history and to build hypotheses \cite{ale2021long}. Before doing so, we state the hypotheses that will be tested:

\subsection*{Hypotheses}
The model describes regime change as a contested transition between dominant shared ideas; political violence is used as an empirical proxy that can reflect such transitions without being identical to them. We therefore compare qualitative features rather than fitting the model to the data. We address the following hypotheses:
\begin{enumerate}
\item[\textbf{H1}] A minimal opinion model that combines cumulative-advantage spreading, a memory that prevents agents from returning to old opinions, and the slow entry of new opinions produces recurrent regime changes, in the sense of contested transitions between dominant opinions, without tuned parameters.
\item[\textbf{H2}] Around a transition, the number of opinion changes $\Psi$ rises approximately exponentially and declines approximately linearly, as spreading is limited by the interface between opinion groups.
\item[\textbf{H3}] Political violence events within a country are temporally clustered, i.e., periods with many events follow each other more often than expected for a random ordering of the same counts.
\item[\textbf{H4}] Political violence, as a proxy for contested transitions, shows the shape of transitions in the model: an approximately exponential rise towards a peak, followed by an approximately linear decline.
\end{enumerate}
H1 and H2 are tested on the model, H3 and H4 on the empirical data.

\section*{Materials and methods}
\subsection*{Model}
Consider $N=L^2$ agents on a two dimensional lattice with periodic boundary conditions, such that every agent has $k$ nearest neighbors. Each agent $i$ is assigned an opinion variable $\sigma_i$ which can assume one integer value, representing an opinion, and will be visualized by a corresponding color. Opinion variables will only be compared to being equal or unequal, which makes the choice of the specific data structure arbitrary.

Opinions spread locally from agent to agent between nearest neighbors, quite as in the classical voter model, however, with one extra constraint: an agent will never accept an opinion that they held before. This constraint is motivated by the human tendency to prefer new over old, as has been used in \cite{bornholdt2011emergence} for modeling the replacement of scientific paradigms. Here we transfer this idea to modeling the turnover of political regimes or dynasties. Even though old ideas can return in history, the general trend is that new ideas take over in the long run. For simplicity we assume that old ideas cannot return.

Further we assume that an opinion spreads with a probability proportional to the number of agents already holding that opinion, motivated by the observed cooperativity in opinion change of humans \cite{asch1951effects,milgram1963behavioral}. This dynamical rule led to ``paradigm shifts'' in \cite{bornholdt2011emergence}, much reminiscent of regime changes in political landscapes, which was an initial motivation for this article. Mathematically, one can account for this with a cumulative advantage, or multiplicative growth, which historically has been formulated with different names \cite{yule1925ii,simon1955class,price1976general,barabasi1999emergence} and in different contexts, including politics \cite{rigney2010matthew,perc2014matthew}. 

A final ingredient to our model is the addition of new opinions to the system: Once in every $N$ micro time steps, an agent is changed to a new opinion that did not yet exist. Note that this happens on a slow time scale, once every sweep for the entire system, reflecting that new opinions are much less likely to emerge as compared to already existing opinions. As the paradigms study \cite{bornholdt2011emergence} has shown that the frequency of ``paradigm shifts'' is largely independent of the emergence rate $\alpha$ of new ideas in that model, we here dropped this parameter altogether by fixing that a new opinion is introduced exactly once per sweep. We thus obtain a model that apart from system size and structure is effectively parameter-free.

In detail, in every macro time step $t \rightarrow t + 1$, or sweep, the following steps are performed: 
\begin{enumerate}
	\item[(i)\thinspace] $N$ times, where $N$ is the number of agents (so that each agent is
  selected once per sweep on average): A pair of a random agent $i$ and one of its $k$ nearest neighbors $j$ is selected. The number $n_j$ of agents with the same opinion as $j$ in the whole system is counted. With probability $p_j = n_j / N$, agent $i$ assumes the opinion of agent $j$ by setting $\sigma_i = \sigma_j$ if and only if agent $i$ never held this opinion in the past.
	\item[(ii)\thinspace] Only once: A random agent $i$ switches to a new opinion that does not yet exist.
\end{enumerate}

For the implementation of the model, each agent $i$ has a blacklist memory of all opinions held in the past. In step (i), agent $i$ compares the opinion proposed by agent $j$ with its blacklist. The simulation is rapidly independent of the initial condition which we here choose such that all agents share the same initial opinion.

\subsection*{Observables}
The number of opinion changes $\Psi(t)$ is the number of accepted opinion changes in step~(i) during sweep $t$. The single agent that switches to a new opinion in step~(ii) is not counted. Since opinions can only change at the boundary between agents of different opinions, $\Psi$ is approximately proportional to the total length of the interfaces between opinion groups.

An observable that characterizes the growing dissent among agents is the fragmentation of opinions in the population, defined as
\begin{equation}
\Phi = \left( \sum_{g=1}^{N_G} p_g^2 \right)^{-1},
\end{equation}
with groups indexed by $g$ and the total number of groups being $N_G$~\cite{laakso1979effective,maslov2015diversity}. As in the model algorithm, the fraction of agents who share the same opinion is $p_g = n_g/N$. For convenience, single agents with a unique opinion are also counted as groups. Intuitively, $\Phi$ is the effective number of groups~\cite{laakso1979effective}.

A transition occurs when an opinion occupies half of all agents for the first time. The period length is the time between two consecutive transitions.

We also record the cumulative fraction of agents that have held a given opinion at least once, which corresponds to the cumulative adoption curve in the diffusion-of-innovations literature.

\subsection*{Simulations}
Unless stated otherwise, simulations use a square lattice with $N = 50^2$ agents, periodic boundary conditions and $k = 4$. To study the dependence on system size, we use $N = 10^2, 50^2, 80^2, 100^2$; to study the dependence on the number of neighbors, we use $k = 4, 8, 12$ and $k = N$ (all agents are neighbors) for $N = 50^2$. Each simulation runs for $10^6$ sweeps.
 
Simulations and analysis were implemented in Python~3.13.5 using NumPy~2.2.6 \cite{harris2020array}, SciPy~1.16.1 \cite{virtanen2020scipy} and scikit-learn 1.7.2 \cite{pedregosa2011sklearn}. Figures were produced with Matplotlib~3.10.5 \cite{hunter2007matplotlib}. All code is available at \url{https://doi.org/10.5281/zenodo.23989457}.

\subsection*{Empirical data}
We use three compilations of events within a state.

\textit{Europe.} Sorokin~\cite{sorokin1937social} compiled internal disturbances for Ancient Greece, Ancient Rome, Byzantium, France, Germany and Austria, England, Italy, Spain, the Netherlands, Russia, and Poland and Lithuania. Internal disturbances are disorders, riots, revolutions and other manifestations of social tension that violated the existing social order and laws of the period; private feudal wars, which were legal at the time, and conflicts of an international character are excluded. All disturbances recorded in standard historical works are included, from small disorders to major revolutions~\cite[pp.~385, 405--407]{sorokin1937social}. Unlike the other two sources, this criterion does not necessarily require violence. We use the average year if a time interval is given for an event.

\textit{United States.} Turchin~\cite{turchin2012dynamics} compiled political violence events in the United States from 1780 to 2010. An event is included only if it caused at least one fatality, and the database focuses on group-level violence. We use the riots, defined as group-on-group violence by groups of twelve or more people asserting their will through force outside the normal bounds of law~\cite{turchin2012dynamics}. The original data are available at \url{https://files.prio.org/Journals/JPR/2012/49/4/Replication\%20Turchin.zip}, while the binned data for the analysis in this paper are also included in the data repository for this work.

\textit{China.} Lee~\cite{lee1931periodic} compiled the frequency of internecine wars in China from 221~B.C. to 1929~A.D. per five-year period, irrespective of the size of a war. From the Qin dynasty to the end of the Ming dynasty, Lee extracted the data largely from Chih Shao-nan's Tih Wang Nien Piao, a chronological table of the reigns of Chinese emperors, and checked them for several periods against the Tung-Kien; for the Qing dynasty he used Japanese chronological tables, and for the period after 1911 largely his own memory~\cite{lee1931periodic}. Internecine wars are uprisings against the ruling authority in times of unity and wars between coexisting rival states within China in times of division; wars against peoples outside the geographic unit of China are excluded. No tabulated version of these data is available. We therefore digitized the values from the lower curve of the figure in~\cite{lee1931periodic} manually with an overlaid grid, since the plotted values are integer counts per five-year interval.

For all series, events are counted in 5-year bins, which is the original resolution of the Chinese data. Counting the number of events instead of, for example, the death toll has the benefit of less sensitivity to event outcomes and a better comparison between societies of different sizes and historical periods. Furthermore, one can expect that if an event was important enough, its existence was recorded, and that the statistics will not be altered substantially if a few events have been missed.

The three compilations differ in what they count: internecine wars for China, lethal riots for the United States, and internal disturbances, including non-violent ones, for Europe. Countries also differ in size and in the time periods of recorded history. We therefore compare the series only through relative quantities within each series (clustering in time and the shape of aligned peaks), never through absolute counts across series.

All event counts, in the binned form used here, are available at \url{https://doi.org/10.5281/zenodo.23989457}.

\subsection*{Statistical analysis}
\textit{Temporal clustering.} For each binned series we compute two statistics. The Fano factor $F = \mathrm{Var}(n)/\langle n \rangle$ of the counts $n$ per bin measures overdispersion relative to a Poisson process with the same mean, for which $F = 1$ as $\mathrm{Var}(n) = \langle n \rangle = \lambda$. It does not depend on the order of the bins. The lag-1 autocorrelation $r_1$, the Pearson correlation between the counts in consecutive bins, measures whether bins with many events follow each other, i.e., clustering in time. To test whether $r_1$ is larger than expected by chance, we randomly permute the bins of each series $10\,000$ times, which keeps the counts but destroys their order, and report as $p$-value the fraction of permuted series with $r_1$ at least as large as observed. A slow change of the event rate over centuries, for example due to better documentation in later periods, also produces a positive $r_1$. We therefore repeat the test after subtracting a centered 100-year (20-bin) moving average from the observed series and from each permuted series ($r_1^{\mathrm{detr}}$, $p^{\mathrm{detr}}$). This isolates clustering on timescales shorter than a century.

\textit{Aligned peaks.} Peaks are identified as local maxima of a series that are above half of the maximal value of that series (\texttt{scipy.signal.find\_peaks}). All peaks of a series are aligned at their maximum, and the counts in a window of $\pm 40$ years (data) and $\pm 100$ sweeps (model) around the peak are summed and divided by the number of peaks. No calibration between sweeps and years is implied; the comparison concerns the shape of the peaks only.
 
\textit{Curve fitting.} For the aligned peaks, we fit an exponential function $f(t) = a\,e^{bt} + c$ to the rising part ($t \le 0$), and both a linear function $f(t) = a t + c$ and an exponential function to the declining part ($t \ge 0$), using nonlinear least squares (\texttt{scipy.optimize.curve\_fit}). We report the coefficient of determination $R^2$ as a measure of goodness of fit (\texttt{sklearn.metrics.r2\_score}). $R^2$ describes how well a function fits the data and is not a significance test. The statistical test for clustering is the permutation test described above.

\section*{Results}
\begin{figure}[!h]\centering
\includegraphics[width=1\linewidth]{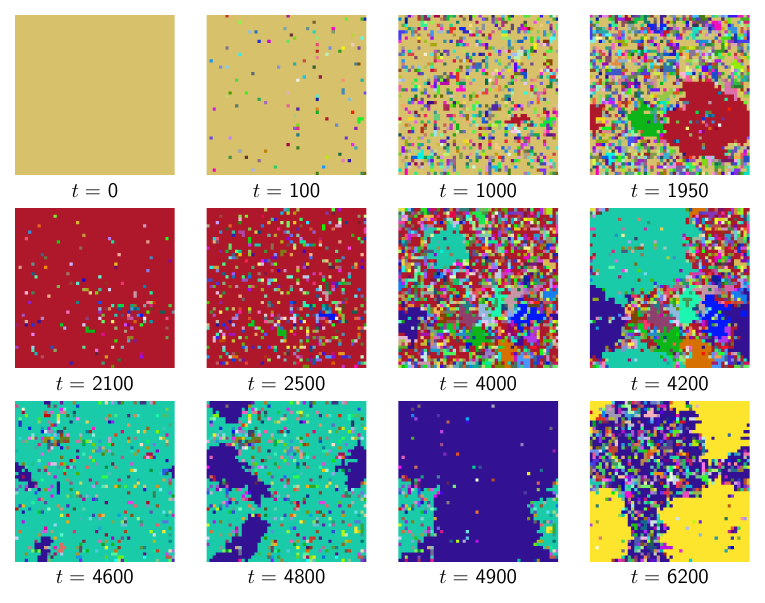}
\caption{\textbf{Simulation example.} Time evolution of a system with $N=50 \times 50$ agents.}
\label{fig:grid_time_evolution}
\end{figure}

\begin{figure}[!h]\centering
\includegraphics[width=1\linewidth]{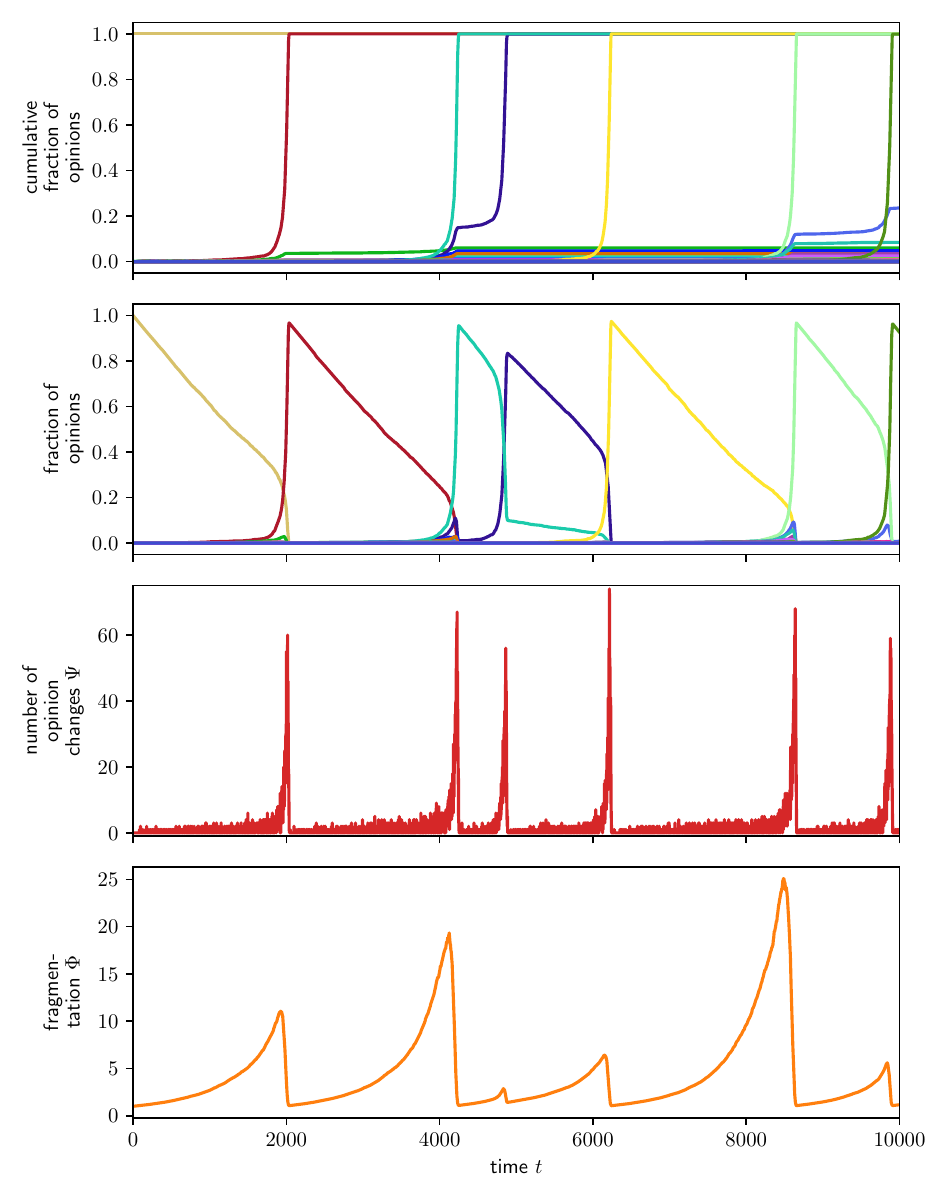}
\caption{\textbf{Example time series.} Statistical measures of the system shown in Fig~\ref{fig:grid_time_evolution}. Compare violence events in Fig~\ref{fig:empirical_data} (red).}
\label{fig:statistics_time_evolution}
\end{figure}

As shown in Figs~\ref{fig:grid_time_evolution} and \ref{fig:statistics_time_evolution}, the fraction of the dominant group slowly decreases as new opinions enter the system. Eventually a new opinion gains a large proportion and takes over the whole system rather quickly. The older group cannot reclaim the agents because of the memory condition of each agent preventing them from accepting old opinions that have been held before.

\begin{figure}[!h]\centering
\includegraphics[width=1\linewidth]{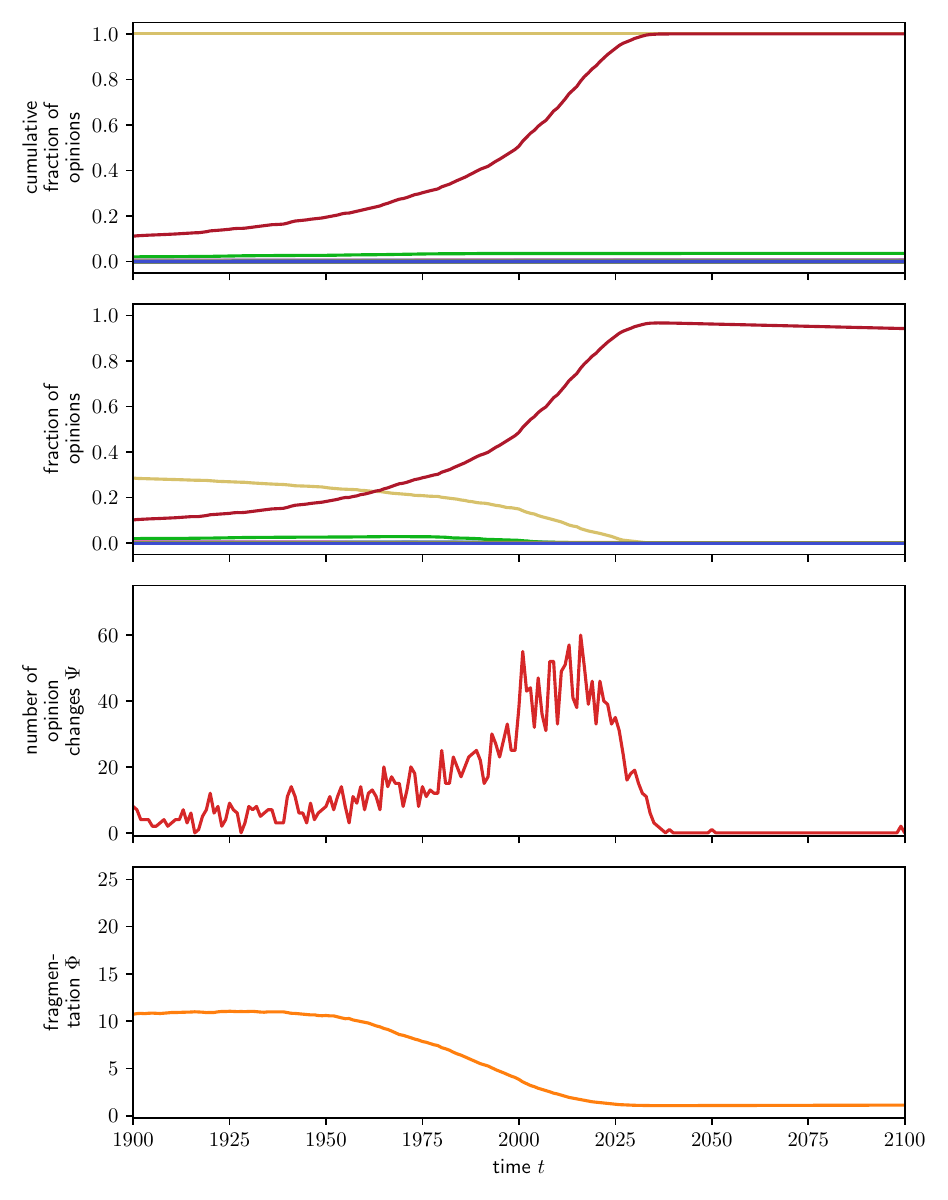}
\caption{\textbf{Zoomed example time series.} Zoomed in time interval for the simulation in Figs~\ref{fig:grid_time_evolution} and \ref{fig:statistics_time_evolution}. Note the approximately linear decrease of opinion changes $\Psi$.}
\label{fig:statistics_time_evolution_zoom}
\end{figure}

Fig~\ref{fig:statistics_time_evolution_zoom} shows a zoomed-in example of a transition from one group to the other. One can see that the number of opinion changes $\Psi$ grows increasingly fast. $\Psi$ peaks at the time when the winning opinion has its largest growth. Afterwards, the new opinion continues to take over the entire system, and $\Psi$ decreases approximately linearly. After this process on a short time scale, the new dominant opinion is slowly eroded by many new opinions on a much longer time scale.

The cumulative fraction of agents that have held the winning opinion (Fig~\ref{fig:statistics_time_evolution_zoom}, top panel) is S-shaped, so the model shows the slowing down at the end of the adoption process known from the diffusion of innovations~\cite{rogers2003,bass1969}. Its form differs from the classical logistic curve for two reasons. First, an opinion spreads through contact and is then accepted with a probability proportional to its overall popularity, so in a well-mixed population the adoption rate of an opinion with population fraction $C$ scales as $C^2(1-C)$ rather than $C(1-C)$, which makes the early spread slower and the takeover more abrupt. Second, interactions are local: since $\Psi$ is approximately proportional to the interface length between opinion groups, the late phase is limited by the shrinking interface rather than by a shrinking pool of non-adopters, and $\Psi$ declines approximately linearly.

Looking at the cumulative fraction of agents that an opinion has reached, one can observe that a dominant opinion will usually reach all agents. This also happens if two opinions are starting to grow large at the same time. In the simulation in Fig~\ref{fig:grid_time_evolution} this can be observed at the end of the second row and the following third row of the figure. An initial takeover of the first opinion is followed by a fast consecutive takeover of the second opinion. 

Peaks can therefore sometimes be close together, when two groups grow large at similar times. Note that this is a different phenomenon compared to so-called father-and-son cycles described by Turchin~\cite{turchin2012dynamics,turchin2016ages}, which are not produced by this model.

\begin{figure}[!h]\centering
\includegraphics[width=1\linewidth]{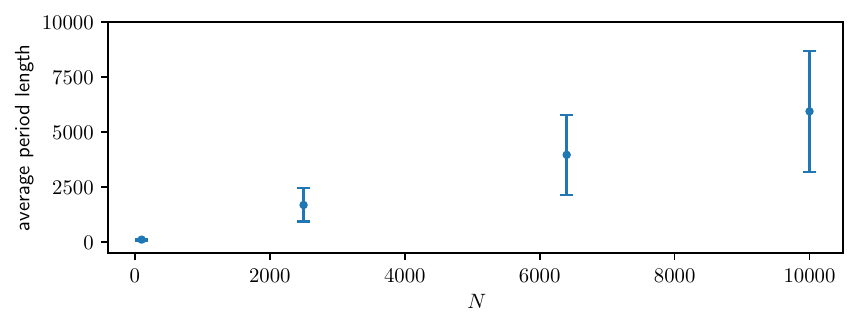}
\caption{\textbf{Period length $N$ dependence.} Average period lengths between transitions of the dominant opinions for system sizes $N=10^2, 50^2, 80^2, 100^2$. Simulations were performed for $10^6$ sweeps. Error bars show the standard deviation (SD) of the period lengths within the simulation. Larger systems exhibit longer periods between transitions.}
\label{fig:average_transition_times_N}
\end{figure}

\begin{figure}[!h]\centering
\includegraphics[width=1\linewidth]{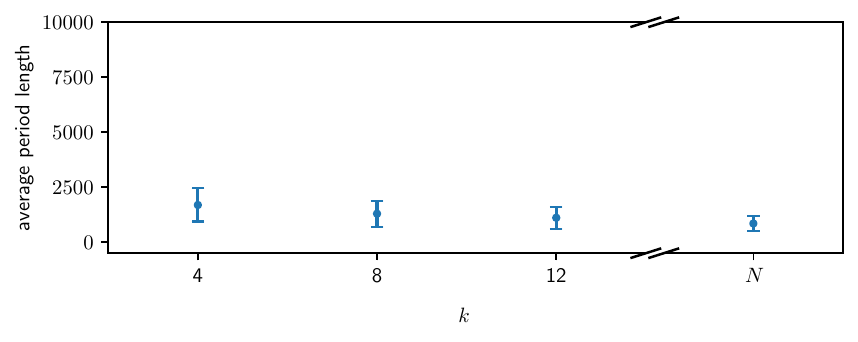}
\caption{\textbf{Period length $k$ dependence.} Average period lengths between transitions of the dominant opinions for $k = 4, 8, 12$ neighbors and the fully connected case $k = N$, for $N = 50^2$. Simulations were performed for $10^6$ sweeps; error bars show the SD of the period lengths within the simulation. As the number of neighbors increases, periods become shorter.} 
\label{fig:average_transition_times_k}
\end{figure}

The only two parameters of the model $N$ and $k$ define the structure in which the model acts. Similarly as in \cite{bornholdt2011emergence}, takeovers of a new dominant opinion occur slower for larger system sizes $N$ as new opinions can only spread through the local neighborhood and the chances of spreading become increasingly small due to the cumulative advantage, which can be observed in Fig~\ref{fig:average_transition_times_N}. Larger coordination numbers $k$ can counteract this behavior, as opinions can spread faster to more distant neighbors, which can be observed in Fig~\ref{fig:average_transition_times_k}. The qualitative patterns in Figs~2--4 are robust to changes in the details of the system. Having new opinions enter the system at different rates as in~\cite{bornholdt2011emergence} does not change the dynamics qualitatively; it is only important that new opinions are added on a slower time scale than the spreading time scale.

\subsection*{Fragmentation as an early warning signal}
Fragmentation $\Phi$ (Materials and methods) grows strongly and increasingly fast before a new opinion takes over the system (Figs~3 and~4). It increases much earlier than the fraction of the new dominant opinion and can thus be used as an early warning measure for transitions in this model. Similarly as in~\cite{horstmeyer2020predicting}, it only uses information on the opinions of the agents, and no structural information is needed. 

\subsection*{Temporal clustering of political violence}
Table~\ref{tab:clustering} quantifies the temporal clustering of political violence events. The lag-1 autocorrelation $r_1$ of the binned counts is positive in all thirteen series, but substantial only for China and the United States ($r_1 \approx 0.65$). In the European series it is weak ($r_1 \leq 0.32$), although often statistically significant because of the large number of bins, and largely disappears after detrending (see below). The Fano factor shows strong overdispersion only for China ($F = 8.56$) and the United States ($F = 19.34$); for the European series it is close to the Poisson value of one.

After removing variation on timescales longer than a century, clustering remains strong in China ($r_1^{\mathrm{detr}} = 0.55$) and the United States ($r_1^{\mathrm{detr}} = 0.38$). In the European series, $r_1^{\mathrm{detr}}$ is small or negative; it is significant at the 5\% level only for France, England and Russia, with values below 0.1.

\begin{table}[!ht]
\centering
\caption{{\bf Temporal clustering of political violence events.} Event counts in 5-year bins (Fig~1). The Fano factor (variance/mean) measures overdispersion relative to a Poisson process with the same mean (Fano factor $=1$). $r_1$ is the lag-1 autocorrelation of the binned counts and measures clustering in time. $p$ is the fraction of 10\,000 randomly shuffled series with a lag-1 autocorrelation at least as large as observed. $r_1^{\mathrm{detr}}$ and $p^{\mathrm{detr}}$ are the same quantities after subtracting a 100-year (20-bin) moving average from the observed and from each shuffled series, removing variation on timescales longer than a century.}
\label{tab:clustering}
\vspace{12pt}
\begin{tabular}{|l|r|r|r|r|r|}
\hline
\textbf{Series} & \textbf{Fano factor} & \textbf{$r_1$} & \textbf{$p$} & \textbf{$r_1^{\mathrm{detr}}$} & \textbf{$p^{\mathrm{detr}}$} \\ \thickhline
China                & 8.56  & 0.650 & $<0.001$ & 0.545  & $<0.001$ \\ \hline
United States        & 19.34 & 0.647 & $<0.001$ & 0.378  & 0.003    \\ \hline
Ancient Greece       & 0.89  & 0.060 & 0.252    & $-0.098$ & 0.680  \\ \hline
Ancient Rome         & 1.21  & 0.319 & $<0.001$ & 0.046  & 0.087    \\ \hline
Byzantium            & 1.04  & 0.109 & 0.068    & 0.049  & 0.110    \\ \hline
France               & 1.27  & 0.264 & $<0.001$ & 0.064  & 0.031    \\ \hline
Germany and Austria  & 1.19  & 0.115 & 0.039    & 0.041  & 0.084    \\ \hline
England              & 1.11  & 0.109 & 0.040    & 0.075  & 0.029    \\ \hline
Italy                & 1.35  & 0.240 & $<0.001$ & $-0.043$ & 0.433  \\ \hline
Spain                & 0.99  & 0.165 & 0.002    & $-0.056$ & 0.523  \\ \hline
The Netherlands      & 1.31  & 0.036 & 0.253    & $-0.120$ & 0.860  \\ \hline
Russia               & 0.95  & 0.231 & $<0.001$ & 0.097  & 0.023    \\ \hline
Poland and Lithuania & 0.90  & 0.123 & 0.056    & 0.022  & 0.188    \\ \hline 
\end{tabular}
\end{table}

\subsection*{Comparison to empirical data: characteristic growth and decline of political violence}
The peaks of opinion changes $\Psi$ resemble qualitative patterns of violence events in the empirical data. In the Chinese and US data, the largest peaks are of similar magnitude, which in the model corresponds to the fast takeover of a new opinion, where the interface length between groups is longest. This might be a sign of a process on a finite limited space that saturates, reminiscent of percolation well above the percolation threshold. The processes involved only take decades in the empirical data.

\begin{figure}[!h]\centering
\includegraphics[width=1\linewidth]{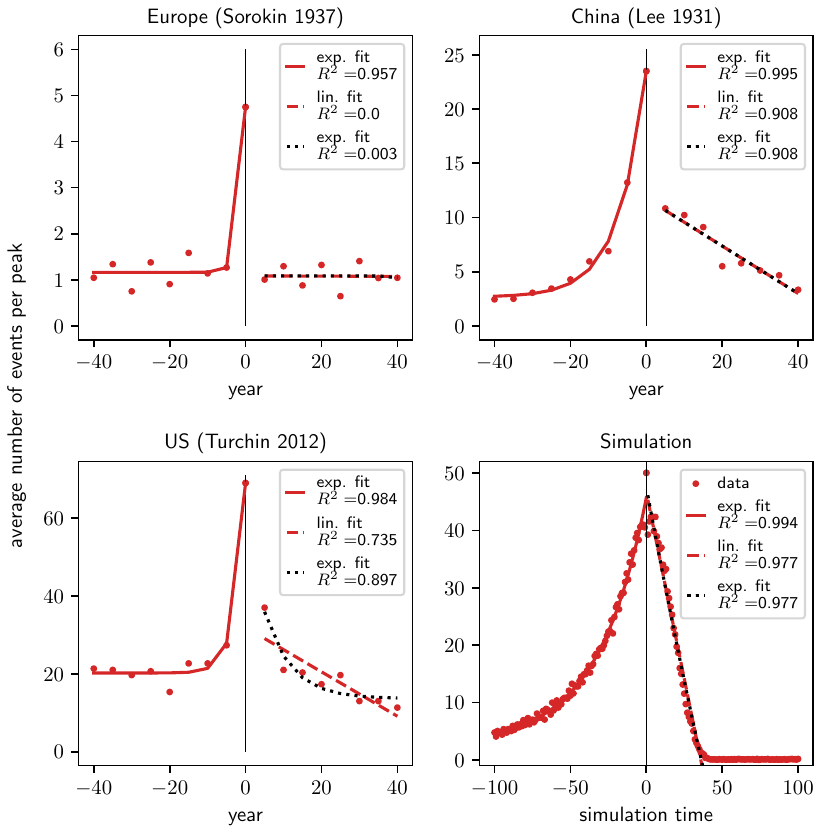}
\caption{\textbf{Characteristic growth and decline.} Aligned and averaged peaks for the datasets in Fig~\ref{fig:empirical_data} and for the simulation in Fig~\ref{fig:statistics_time_evolution}. Negative times are before the peak, positive times after it. Lines show least-squares fits of an exponential function to the rise and of a linear (dashed) and an exponential (dotted) function to the decline, with coefficients of determination $R^2$. No calibration between sweeps and years is implied; see Materials and methods for peak detection and fitting.}
\label{fig:aligned_peaks}
\end{figure}

Aligning all peaks of a series (Fig~\ref{fig:aligned_peaks}) shows the characteristic growth and decline. For the Chinese and US data, the rise towards the peak accelerates and is well described by an exponential function ($R^2 = 0.995$ and $0.984$), as is the rise of $\Psi$ in the model, which arises from the cumulative advantage in model step~(i). For China, the decline after the peak is well described by a linear function ($R^2 = 0.908$),  and an exponential function does not improve the fit. This is consistent with the approximately linear decline of $\Psi$ in the model. For the United States, an exponential function describes the decline better than a linear one ($R^2 = 0.897$ versus $0.735$).

The European data~\cite{sorokin1937social} are too sparse for a meaningful fit, consistent with the weak clustering in Table~\ref{tab:clustering}. Note that most historical countries in the European data are also much smaller than the historical Chinese Empire and the United States.

Peaks just a few decades apart, as produced occasionally by the model, can also be observed in a few cases in the empirical data. Turchin gives arguments for this behavior in the sense that conflicts are not properly settled and reappear faster~\cite{turchin2016ages}; here the phenomenon arises purely from the model dynamics.

\section*{Discussion}
In a radically simple, parameter-free opinion formation model, we have modeled the eternal renewal of political opinions. A remarkable observation is that the sizes of the actively moving interfaces, where new opinions replace the old, resemble qualitative features of the growth and decline of political violence in history, which in the Chinese and US data also shows peaks of similar magnitude. Characteristic growth and decline can be argued to arise from a cumulative advantage that saturates, resulting in peaks of similar magnitude.

The three data compilations differ in what they count: internecine wars for China, lethal riots for the United States, and internal disturbances for Europe. Countries also differ in size, and the length of the recorded history differs between countries. These differences affect absolute numbers of events, which is why we do not compare counts across series. Both quantities we use, clustering in time and the shape of aligned peaks, are relative quantities within a series. They are affected by these differences only if the recording criteria changed within a series during a peak, which we consider unlikely on timescales of decades. A slow change of documentation over centuries, by contrast, can affect clustering statistics, which is why we report them also after removing variation on timescales longer than a century (Table~\ref{tab:clustering}).

We used violent events as a proxy for contested transitions, not for regime change itself. The link between the two is not one-to-one: regimes have changed peacefully, and periods of violence have ended without a change of regime. In the model, $\Psi$ measures the activity at the interfaces between competing opinions, regardless of the outcome. A peak of $\Psi$ is always followed by a new dominant opinion in the model, whereas in history a contest may also be resolved in favor of the existing order. Data that distinguish violent episodes with and without a subsequent change of regime would allow a sharper test of the model. The aligned peaks provide an empirical pattern that can be linked to models as described in~\cite{sobkowicz2009modelling,flache2017models}. However, connecting the model behavior to empirical data remains difficult because the data are sparse.

Independent of the comparison with data, a central property of the model is that it has no free parameters. As the system size and neighbors are part of the system structure, the only processes influencing the dynamics are that new opinions are added and that they spread in proportion to their size as a cumulative advantage. The important element is that the two processes act on different time scales. This is an essential property of the system and does not need any finely tuned parameter.

Besides the interpretation in terms of elites introduced above, the spatial structure of the model also allows a geographical interpretation, as in~\cite{guzman2018ecosystems}. The geographical size of a country would then act as a finite limiting space on which violence spreads, and the interface length between two groups in the model is reminiscent of the line of contact between two groups on a geographical space, as, for example, in a civil war. Since the interface length is approximately proportional to the number of opinion changes $\Psi$, this connects the model directly to the violence patterns in the empirical data. In pre-industrial times, communication was mostly local, such that a two-dimensional spread seems a realistic assumption. Both interpretations are not exclusive, as elite networks in pre-industrial societies were largely local as well.

In summary, the model studied here connects opinion dynamics and patterns of political violence in a simple but intuitive way. Several qualitative patterns observed in the empirical data can be observed in the model: recurrent transitions, peaks of similar magnitude, and an accelerating rise towards a peak followed by a decline, which in the Chinese data is approximately linear as in the model. As peaks of violence appear with similar magnitudes, one can argue for a saturation effect of political violence events on a finite space. This might contribute to a better understanding of the political conflicts that shaped history in the long run. The model shows how a cumulative advantage that saturates might affect the recurring characteristic growth and decline of political violence.

\bibliography{literature}

\end{document}